\documentclass[aps,showpacs,prc,preprint,superscriptaddress,floatfix,nofootinbib]{revtex4}
\usepackage{amsfonts}
\usepackage{ulem}
\usepackage{xcolor}
\usepackage{amssymb}
\usepackage{amsfonts}
\usepackage{amsmath}
\usepackage{amssymb}
\usepackage{amsthm}
\usepackage{epsf}
\usepackage[dvips]{graphicx}

\def\bk{{\mbox{\boldmath$k$}}}
\def\bq{{\mbox{\boldmath$q$}}}
\def\bp{{\mbox{\boldmath$p$}}}
\def\bgam{{\mbox{\boldmath$\gamma$}}}

\def\calP{{\cal P}}
\def\pk{\dfrac{\bp\bk}{\bp^2}}
\def\qk{\dfrac{\bq\bk}{\bq^2}}
\def\la{\langle}
\def\ra{\rangle}
\begin{document}
\title{Extending the truncated Dyson-Schwinger equation to finite temperatures}
 \author {S.~M. Dorkin}
\affiliation{Bogoliubov Lab.~Theor.~Phys., 141980, JINR, Dubna,  Russia}
\affiliation{International University Dubna, Dubna, Russia }
\author { M.~Viebach\footnote{Now at Institute of Power Engineering, TU Dresden, 01062 Dresden, Germany}}
\affiliation{Institut f\"ur Theoretische Physik, TU Dresden, 01062 Dresden, Germany}
 \author{L.~P. Kaptari}
 \affiliation{Bogoliubov Lab.~Theor.~Phys., 141980, JINR, Dubna, Russia}
\affiliation{Helmholtz-Zentrum Dresden-Rossendorf, PF 510119, 01314 Dresden, Germany}
\author { B.~K\"ampfer}
\affiliation{Helmholtz-Zentrum Dresden-Rossendorf, PF 510119, 01314 Dresden, Germany}
\affiliation{Institut f\"ur Theoretische Physik, TU Dresden, 01062 Dresden, Germany}
    \begin{abstract}
  In view of the properties of  mesons  in hot strongly interacting matter the properties of the solutions of the
truncated Dyson-Schwinger equation for the quark propagator at finite temperatures within the rainbow-ladder
approximation are analysed in some detail. In Euclidean space within the Matsubara imaginary time
formalism the quark propagator is   not longer  a O(4) symmetric function and possesses
a discrete  spectrum of the fourth component of the momentum. This makes the treatment of the Dyson-Schwinger
and Bethe-Salpeter equations conceptually different from the vacuum
and technically much more involved.  The   question whether the interaction kernel known from
vacuum calculations can be applied at finite temperatures remains still open. We find that, at low
temperatures, the model interaction with vacuum parameters provides a reasonable description of the
quark propagator, while at temperatures higher than a certain critical value $T_c$ the interaction requires stringent
modifications. The general properties of the quark propagator at finite temperatures can be inferred from
lattice QCD   (LQCD)   calculations. We argue that, to achieve a reasonable agreement of the model calculations
with that from LQCD, the kernel is to be modified in such a way as to screen the infra-red part
of the interaction at temperatures larger than $T_c$. For this, we analyse the solutions of the
truncated Dyson-Schwinger equation
with existing interaction kernels in a large temperature range with particular attention on
high temperatures in order to find hints to an adequate temperature dependence of the interaction kernel to be further
implemented in   the Bethe-Salpeter equation for mesons. This will allow to investigate the possible
in medium modifications of the meson properties
as well as the conditions of  quark deconfinement in hot matter.
\end{abstract}
 \maketitle
\section{Introduction}
     The description of mesons as quark-antiquark bound states within the framework of the Bethe-Salpeter (BS) equation
 with momentum dependent quark mass functions, determined by the Dyson-Schwinger (DS) equation, is able
 to explain successfully many spectroscopic data, such as meson masses
 \cite{rob-1,MT,ourLast,Hilger,Maris:2003vk,Holl:2004fr,Blank:2011ha},   electromagnetic properties
 of pseudoscalar mesons and their radial excitations~\cite{Jarecke:2002xd,Krassnigg:2004if,Roberts:2007jh}
  and  other   observables~\cite{ourFB,wilson,JM-1,JM-2,rob-2,Alkofer,fisher,Roberts:2007jh}.
 Contrary to purely  phenomenological models, like  the quark bag  model,
 such a formalism  maintains   important features of QCD, such as dynamical chiral
 symmetry breaking, dynamical quark dressing, requirements of the renormalization group theory etc., cf.~Ref.~\cite{physRep}.
 The main ingredients here are  the full quark-gluon vertex function and  the dressed gluon
 propagator,  which are entirely determined  by  the running coupling
 and the bare quark mass parameters. In principle, if one were able to solve the Dyson-Schwinger equation,
 the approach would  not depend on  any additional parameters.
   However, due to  known technical problems,
  one restricts oneself to calculations within effective models which
  specify the dressed vertex function $\Gamma_\mu$ and interaction kernel $D_{\mu\nu}$.
  The rainbow-ladder approximation \cite{MT} is a model with rainbow truncation of the
  vertex function $\Gamma_\mu\to \gamma_\mu$ in the quark DS equation  and
  a specification of the dressed quark-quark interaction kernel
   as $ g^2 D_{\mu\nu}(k) \to {\cal G}(k^2) D_{\mu\nu}^{free}(k)$. (Here, $\gamma_\mu$
   is a Dirac gamma matrix and $D_{\mu\nu}$ stands for the gluon
   propagator; $g$ is the coupling strength and $k$ denotes a momentum.)

   The model is completely specified once a form
is chosen for the “effective coupling” ${\cal G}(k^2)$. The ultraviolet
behavior is chosen to be that of the
QCD running coupling $\alpha(k^2)$; the ladder-rainbow
truncation then generates the correct perturbative
QCD structure of the DS and BS
equations. Moreover, the  ladder-rainbow truncation preserves such an  important feature of the
 theory as the maintenance of the Nambu-Goldstone theorem in the chiral limit,
according to which   the spontaneous chiral symmetry breaking results in an appearance of a
 (otherwise absent) scalar term in the quark propagator of the DS  equation. As a consequence, in the BS
 equation a massless pseudoscalar bound state should appear . By using the Ward  identities, it has been proven (see,
 e.g.  Refs.~\cite{smekal, scadron,munczek}) that in the chiral limit the DS equation for the quark
 propagator and     the BS equation for a massless pseudo-scalar in ladder approximation are completely  equivalent.
 It  implies  that such a  massless bound state (pion) can be interpreted as a
    Goldstone boson. This results in a straightforward understanding    of the pion as both a
     Goldstone boson and  quark-antiquark bound state.

    Another important property of the DS and BS equations is their explicit Poincar\'{e}  invariance.
    This  frame-independency of the approach provides a
    useful tool in studying processes  when  a rest frame for mesons cannot or
    needs not be defined.

 The merit of the approach is that, once the effective parameters are fixed
 (usually the effective parameters of the
 kernel are chosen, cf. Ref.~\cite{Parameterslattice,williams},  to reproduce the known
 data from lattice calculations, such as the quark mass function and/or  quark condensate),
 the whole spectrum of known mesons is supposed to  be described,   on the same footing,
 including also excited states. The achieved  amazingly good description  of
 the mass spectrum with only  few effective parameters encourages one to employ the same approximations
 to the   truncated Dyson-Schwinger (tDS)  and truncated Bethe-Salpeter (tBS)  equations also at finite temperatures with the hope that, once an adequate
 description of the quark propagators at non-zero temperature  ($T$)  is accomplished, the corresponding solution
 can be implemented in to the BS equation for mesons  to investigate the meson properties
 in hot and dense matter.

   At low temperatures the properties of hadrons   in nuclear matter
   are expected to change in comparison with the vacuum ones, however the main  quantum
   numbers, such as  spin and orbital momenta, space and inner parities etc. are maintained.
   The hot environment may modify the hadron masses, life time  (decay constant) etc. Contrarily,
   at sufficiently large temperature in hot and dense strongly interacting matter,
     phase transitions may occur, related to quark deconfinement phenomena, as e.g.
   dissociation of hadrons in to quark degrees of freedom. Therefore, these temperature regions are of great interest,
   both from a theoretical and experimental point of view. Hitherto,   the truncated DS and BS formalism has   been mostly used
    at large   temperatures  to investigate the critical phenomena
    near and above the  pseudo-critical and  (phase) transition values predicted by lattice simulation data
   (cf. Refs.~\cite{kitaizyRpberts,robLast,MTrenorm,rischke} and  references  therein quoted).
  It has been found that, in order to achieve an agreement of   the  model results with lattice data,
  a modification of the vacuum interaction kernel is required.
   Namely, the infra-red term has to vanish abruptly in this region.
   Accordingly, it has been  suggested~ \cite{kitaizyRpberts,robLast}  to employ
    a kernel with a  Heavyside step-like behaviour  in the vicinity
   of the (pseudo-) critical temperature  $T_c$. Then,  it becomes possible to  achieve a
  rather reliable description of such quantities as the quark spectral function,
  plasmino modes, thermal masses etc., see also~Ref.~\cite{kitazavaPRD80}. However, a use   of  such a
   discontinuously  modified interaction in
  the BS equation in the whole temperature range becomes hindered. Another strategy of solving the DS equation in a
  larger interval of temperatures  is to  utilize directly   the available LQCD results   to fit,
  point by point, the interaction kernel
  at given  temperatures. In such a way one achieves a good description
  of the quark mass function and condensate  for  different   temperatures,  including the region
  beyond  $T_c$~\cite{fischerPRD90,FischerRenorm}. The success of such approaches demonstrates that the
  rainbow approximation to the DS equation with a proper choice of the interaction kernel  is quite
  adequate in  understanding the properties  of quarks in   hot environment. Nevertheless, for   systematic studies of quarks
  and hadrons within the BS equation, on needs a  smooth parametrization of the  kernel  in the whole interval of the considered
   temperatures. In view of still scarce LQCD data, such a direct parametrization from "experimental" data is problematic.
   An alternative method  is to solve simultaneously   a (truncated) set of Dyson-Schwinger equations
  for the quark and gluon propagators within some additional approximations~\cite{fischerGluon}.
  This approach also provides
  good description of quarks in vicinity of $T_c$, however it becomes too cumbersome in attempts
  to solve the BS equation,  since in this case one should solve a too large system of equations.
It should also be  noted that  there are other  investigations of the quark propagator  within the rainbow truncated   DS equation,
which   employ solely the vacuum parameters in  calculations of  $T$-dependencies of quarks ~\cite{BlankKrass}
   without further attempts to accommodate  the kernel   to LQCD results. As a result    one
  finds that  the  critical behaviour of the propagators (e.g.  chiral symmetry restoration)  starts at temperatures much smaller than the ones  expected from  LQCD.

 In the present paper we are interested in a detailed   investigation of the quark propagator in the whole
 range of temperatures, from
 zero   temperatures up to above $T_c$, and find a  reliable smooth  parametrization of the
 kernel. We start with the interaction kernel known at $T=0$ and extend it, step by step,
 to larger temperatures by finding the prerequisites  to meet the requirements of the LQCD and to  be  able to
 implement the kernel into the BS equation in subsequent   studies of the hadron bound states at finite temperatures.

 In quantum field theory, a system embedded  in a heat bath can be
 described within the  imaginary-time formalism,
 known also as the Matsubara approach~\cite{matzubara,kapusta,abrikosov}. Due to finiteness of the heat bath
 temperature $T$ the Fourier transform  to  Euclidean momentum space becomes discrete, resulting
 in a discrete spectrum of the energy, known as the Matsubara frequencies. Consequently, the
 interaction kernel and the DS solution become also discrete with respect to these frequencies.
 Moreover, since the heat bath already fixes a particular frame, the corresponding DS and BS equations
 are not longer $O(4)$ symmetric. This requires a separate treatment of the transversal and longitudinal
 parts of the kernel with the need of an additional function in parametrizing  the quark propagators.
 All this makes the consideration of the DS equation different from the vacuum case. However,
 here is the hope that the phenomenological interaction kernel defined at $T=0$ can be, to some extent,
 applied for finite temperatures as well.

In the present paper we investigate  the prerequisites to the interaction
kernel of the  DS   formalism at finite temperatures to be able
to investigate, in a subsequent step, different processes   with the challenging  problem of
 changes of meson characteristics at finite temperatures.
 Our goal is to determine with what extend the rainbow truncation of the
 DS equation is applicable in a large interval of temperature, starting from   low values,
 with the effective parameters, known to
accomplish an excellent description of the hadron properties in vacuum, towards temperatures  above the
 critical values predicted by lattice calculations. We try to find a proper modification of the kernel
  at higher temperatures to be able to describe the properties of the quark propagator in the whole
  temperature range.
 A reliable parametrization   of the $T$-dependence will allow to  implemented it directly into
 the BS equation in the same manner as at  $T=0$ and to investigate, e.g. in-medium changes of mesons in  hot environment.
This is  crucial, e.g.  in understanding the di-lepton yields in nucleus-nucleus collisions.
Our future goal is to investigate to what extend the effective parameters, known to
accomplish an excellent description of the hadron properties in vacuum, can be utilized in the
 BS equation to investigate the hadron modifications  in hot and dense matter
 below and above the critical or cross-over temperature. For this we consider the quark propagators from the DS equation
 in a large temperature range  and investigate their properties and compare qualitatively
 with other approaches, such as the   LQCD calculations.

 Our paper is organized as follows.
 In Sec.~\ref{s:bse},  we recall  the truncated  BS and DS
 equations in vacuum and at finite temperatures.
  The rainbow approximation for the DS equation kernel in vacuum is  introduced
  and  the  system of equations for the quark propagator, to be solved at finite temperature,
 is presented. Numerical solution for the chirally symmetric case
 is discussed in Sec.~\ref{sec3}, where the chiral quark condensate
 and spectral representation for the quark propagator are introduced. It is found that, to achieve a reasonable
 behaviour of the spectral functions above the critical temperature, a modification of the interaction kernel is needed.
 In Sec.~\ref{sec4} we consider the solution of the truncated DS   equation  for finite bare masses. The inflection points of the quark
 condensate  and the mass function are considered as a definition of the pseudo-critical temperature
 at finite quark masses. The procedure of regularization of integrals in calculating the quark condensate
 from the solution of the  DS equation is discussed in some detail.  It is shown that, for finite quark masses,
 the inflection method determines the pseudo critical temperatures by $\sim 50\%$ smaller than the ones
 obtained by other approaches, e.g. by lattice QCD calculations. The possibility to reconcile
 the model and lattice QCD results is considered too.
The impact of the infrared term in the interaction kernel in the vicinity and above the critical
temperature  is also briefly discussed.  Summary and conclusions
are collected in Sec.~\ref{summary}.    A  brief   explanation of the meaning of the rainbow-ladder approxiamtion
is presented in  the Appendix.

\section{ Basic Formulae}
\label{s:bse}

 \noindent
 \subsection{Dyson-Schwinger  and Bethe-Salpeter    equations in vacuum}
\label{Bet}
To determine the bound-state mass of a quark-antiquark pair one needs to solve the
DS and the {homogeneous}   BS equations, which
in the rainbow ladder approximation    and in Euclidean space  read
\begin{eqnarray}&&
S^{-1}(p)= S_0^{-1}(p) + \frac 43 \int \frac
{d^4 k }{(2\pi)^4} \left[g^2 {D}_{\mu \nu}(p-k) \right]\gamma_{\mu} S(k)
\gamma_{\nu}\: ,
\label{sde}\\ &&
    \Gamma(P,p) =   -\frac 43  \int \frac {d^4k}{(2\pi)^4}
    \gamma_{\mu} S(\eta_1 P+k) \Gamma(P,k) S(-\eta_2 P+k))\gamma_{\nu}
    \left [g^2    { D}_{\mu \nu}(p-k)\right ] \: ,
\label{bse}
\end{eqnarray}
where $\eta_1$ and $\eta_2=1-\eta_1$ are the partitioning parameters  defining the quark momenta
as $p_{1,2}=k \pm \eta_{1,2} P $ with $P$ and $k$ denoting the total and relative momenta
of the bound system, respectively;\footnote{
  {Usually,  for quarks of masses $m_{1,2}$ the partitioning
parameters  are chosen as  $\eta_{1,2}=m_{1,2}/({m_1+m_2})$. However, in general the BS solution
is independent of the choice of $\eta_{1,2}$.    }}
 $\Gamma(P,k)$ stands for  the BS vertex function being a $4\times 4$ matrix,
$S_0(p)=\left (i\gamma\cdot p +m\right)^{-1}$  and
$S(p)=\left ( i\gamma\cdot p A(p)  +B(p)\right)^{-1}$ are the
   propagators of bare and dressed  quarks, respectively with mass parameter $m$ and
   the dressing functions  $A(p)$ and $B(p)$. {In  Euclidean space
    we use the  Hermitian matrices  $\gamma_4=\gamma_0,\bgam_{E}=-i\bgam_{M}$ which obey
    the anti-commutation relation $\{\gamma_\mu,\gamma_\nu\}=2\delta_{\mu,\nu}$; for
    the four-product
    one has $(a\cdot b)=\sum_{i=1}^4 a_i b_i$.
          The masses $M$  of mesons as bound states  of a $m_1$-quark and $m_2$-antiquark follow
   from the  solution of BS equation, $P^2=M^2$, in specific $J^{PC}$ channels,
   with the solution of the DS equation (\ref{sde}) as input into the calculations in
   Eq.~(\ref{bse}). The interaction between  quarks in the  pair is encoded in  $g^2 D_{\mu\nu}$
    imagined     as gluon exchange. For consistency, the same interaction is to be employed
     in the DS equation~(\ref{sde}) for the inverse dressed quark propagator.

Often, the coupled equations of the quark propagator $S$, the gluon propagator $D_{\mu\nu}$ and the
quark-gluon vertex
function $\Gamma_\mu$, all with full dressing (and, if needed, supplemented by
ghosts and their respective vertices), are  considered as an
integral formulation being equivalent to QCD.
 {In practice, due to  numerical problems, the finding of the  exact solution of the
system of coupled equations for  $S-D_{\mu\nu}-\Gamma_\mu$ can hardly be accomplished  and
therefore some approximations~\cite{fisher,Maris:2003vk,physRep} are appropriate.
 Note that we must know the form
of $D_{\mu\nu}(k)$ and $\Gamma_\mu (k,p)$, not only in the ultraviolet range, where
perturbation theory is applicable, but also in the infra-red range,
where perturbation theory fails and lattice simulations are   to be corrected for
 finite-volume effects. $D_{\mu\nu}(k)$ and $\Gamma_\mu)(k,p)$  satisfy
DS equations. However, studies of these equations in QCD are
rudimentary and are presently  used only to suggest
qualitatively reliable \textit{ans\"atze} for these   functions.
That is why the quantitative studies of the quark DS equation to date
have employed model forms of the gluon propagator and quark-gluon vertex.
Leaving  a detailed discussion of the variety of approaches in dressing
 of the gluon propagator and vertex function in DS equations
 (see e.g.~Refs.~\cite{Fischer:2008uz,PenningtonUV} and references therein
 quoted) we mention only that in solving the DS equation for the quark propagator
 one usually employs truncations of the
 exact interactions and replaces the gluon propagator combined with the vertex
 function by an effective interaction kernel $[g^2 D_{\mu\nu}]$.
 This leads to the truncated Dyson-Schwinger  equation  for
 the quark propagator which may be referred   to as the gap equation.
In explorative  calculations, the choice of the form of the effective interaction is inspired by
results from calculations of Feynman diagrams within  pQCD maintaining
requirements of symmetry and asymptotic behaviour  already implemented,
cf.~Refs.~\cite{Maris:2003vk,Roberts:2007jh,physRep,PenningtonUV}.
 The  results of such calculations, even in the simplest case of accounting only for one-loop
 diagrams with proper regularization
 and renormalization procedures,  are rather cumbersome for further use in
 numerical calculations, e.g. in the framework of BS or Faddeev equations.
 Consequently, for practical purposes,  the wanted exact results are
 replaced by  suitable parametrizations of the vertex and the gluon propagator.
 Often, one employs  an  approximation which corresponds to one-loop calculations of diagrams with
 the  full  vertex function $\Gamma_\nu$,
 substituted by the free one,
  $\Gamma_\nu(p,k)\rightarrow\gamma_\nu$ (we suppress the color structure and
  account cumulatively for the strong coupling later on).
   To emphasize
the replacement of combined gluon propagator and vertex we use the notation
$[g^2D_{\mu\nu}]$, where an additional power of $g$ from the second undressed vertex is
included.

\subsection{Choosing an interaction kernel}\label{choos}

 Note  that the nonperturbative behaviour of the kernel $[g^2 D_{\mu\nu}]$
 at small momenta, i.e. in the infra-red (IR) region, nowadays is
 not uniquely determined and, consequently, suitable models are needed.
 In principle, constraints on the infra-red form of the kernel  can be sought from studies of the DS equations
with  the fully dressed gluon propagator, $D_{\mu\nu}(k)$, and the dressed gluon-quark vertex $ \Gamma_\nu (p,k)$.
However, there is almost no information available from DS equation studies;
the gluon propagator itself has been often studied via the gap equation,  and  from such
studies one can  merely qualitatively conclude that the gluon propagator  is  enhanced in the infra-red.  There are several \textit{ans\"atze} in the literature for the IR  kernel, which can be
formally classified in the two groups: (i) the IR part is parametrized  by  two
terms - a delta distribution at zero momenta and an exponential, i.e. Gaussian term,
and (ii) only the Gaussian term is considered.
 In principle, the IR term must be supplemented
by a ultraviolet (UV) one, which assures the correct asymptotics at large momenta.
A detailed investigation~\cite{souglasInfrared,Blank:2011ha} of
 the interplay of these two terms has shown that, for bound states,
 the IR part  is dominant for light ($u$,  $d$ and $s$)  quarks
 with a decreasing role for heavier ($c$ and $b$) quark  masses  for which the
 UV part may be quite important in forming  mesons with masses $M_{q\bar q}> 3-4$ GeV as bound states.
 In the vacuum, if one is interested in an analysis of light mesons  with $M_{q\bar q}\le 3-4$ GeV, the UV term can be omitted.
 This is not the case for finite temperatures ($T$)  where one can expect that
  at sufficiently large $T$  some  phase transition  can occur  and/or  quark dissociation of mesons
  into quark degrees of freedom in hot matter. In such a temperature range,
 the IR term  is expected to be screened~\cite{robLast,kitaizyRpberts} and, consequently,  the perturbative UV behaviour can become
 important even for light mesons.

Following examples in the literature \cite{Alkofer,MT,Maris:2003vk,Krassnigg:2004if,wilson,Roberts:2007jh}
the interaction kernel in the rainbow approximation in the Landau gauge is chosen
as
\begin{eqnarray}&&
 g^2(k^2) {\cal D}_{\mu \nu} (k^2) =\left(
 D_{IR}(k^2)+ D_{UV}(k^2)\right) \left( \delta_{\mu\nu}-\frac {k_{\mu} k_{\nu}}{k^2} \right), \nonumber \\ &&
D_{IR}(k^2)=
        \frac{4\pi^2 D k^2}{\omega^6} e^{-k^2/\omega^2},
\ \quad
 D_{UV}(k^2) =
         \frac {8\pi^2 \gamma_m F(k^2)}{
            \ln[\tau+(1+\frac{k^2}{\Lambda_{QCD}^2})^2]} ,
\label{phenvf}
\end{eqnarray}
 where the first term originates from  the effective IR part of the interaction
 determined by soft, non-perturbative effects, while the second one ensures the correct
UV asymptotic behaviour of the QCD running coupling.
 In what follows we restrict ourselves to  two
 models. (i) The interaction consists of both the IR and UV terms: Such an
  interaction  is known as the Maris-Tandy (MT) model~\cite{MT}.
(ii) The UV term is ignored at all: This interaction
 is known as Alkofer-Watson-Weigel \cite{Alkofer} kernel, referred  to as the
  AWW model.
     It should be noted
that at zero temperatures
these models, with  only a few adjustable parameters -   the IR strength   $D$,     the slope parameter    $\omega$ and  quark mass parameter $m$ in the AWW
model and additionally $\tau$, $\Lambda_{QCD}$, $\gamma_m$ and $m_t$ in the formfactor
$ F(k^2)=\left (1-\exp\{-k^2/[4m_t^2]\}\right )/k^2$ in the MT model -
provide a  good description of the pseudoscalar, vector and tensor meson   mass spectra
\cite{Hilger,Holl:2004fr,Blank:2011ha,ourLast}. Therefore, at finite temperatures
a tempting choice of the interactions  is to keep them the same as in vacuum.
 \subsection{Finite temperatures}\label{finite}
 The theoretical treatment of  systems at non-zero temperatures differs from the
 case of zero temperatures.   {In this case,} a preferred  frame is determined
 by the local rest system of the  thermal bath. This means that the $O(4)$ symmetry
 is broken and,  consequently, the
   dependence of the quark propagator on $\bf p$ and $p_0$  requires a separate treatment.
 To describe the propagator in this case a third function $C$ is needed,   besides the functions $A$ and $B$
 introduced above for vacuum.
 { Yet, the theoretical formulation of the field theory
at finite temperatures can be performed in at least two, quite different, frameworks which
treat  fields  either with ordinary time variable
$t  \, ( - \infty  < t < \infty)$, e.g.  the termo-field dynamics
(cf.~\cite{umezawa}) and path-integral formalism (cf.~\cite{nemi,landshof}),
or with imaginary time $it= \tau$ ($0< \tau< 1/T$) which is known as the Matsubara
formalism~\cite{matzubara,kapusta,abrikosov,landsman}.
  In this paper we utilize the imaginary-time formalism within which the
partition function is defined and  all calculations are performed in Euclidean space.}
 Since at $T\ne 0$ the (imaginary) time evolution is restricted to the interval $[0\ldots 1/T]$,
 the quark  fields become    {anti-periodic}
in time with the period $1/T$. In such a case the Fourier transform
is not longer continuous and the energies $p_4$ of particles become discrete
  \cite{matzubara,kapusta,abrikosov} which are known as the Matsubara frequencies, i.e.
  $p_4=\omega_n=\pi T (2n+1)$ for Fermions ($n$ is an integer, running from minus   to plus infinity).
 The inverse quark propagator is now parametrized as
\begin{eqnarray}
 S^{-1}({\bf p},\omega_n)=i\bgam\bp A(\bp^2,\omega_n^2)+i\gamma_4 \omega_n C(\bp^2,\omega_n^2)+
B(\bp^2,\omega_n^2).
\label{inversProp}
\end{eqnarray}
Accordingly, the interaction kernel  is decomposed in to a transversal and longitudinal part
\begin{eqnarray}
[g^2 D_{\mu\nu} (\bq,\Omega_{mn})] =
\calP^T_{\mu\nu} D^T(\bq,\Omega_{mn},0)+
\calP^L_{\mu\nu} D^L(\bq,\Omega_{mn},m_g),
\label{kernel}
\end{eqnarray}
where $\Omega_{mn}=\omega_m-\omega_n$
and the gluon screening (Debye) mass $m_g$ is introduced in the longitudinal part of the
propagator, where $q^2=\bq^2+\Omega_{mn}^2+m_g^2$ enters.
The scalar coefficients $D^{L,T}$ are defined below.
The  projection operators $\calP_{\mu\nu}^{L,T}$ can be written as
\begin{eqnarray}&&
\calP ^T_{\mu\nu}=\left\{  \begin{array}{llll}
                                         0,&&\mu, \nu=4,  \\
                                         \delta_{\alpha\beta}-\dfrac{q_\alpha q_\beta}{\bq^2};&&\mu,\nu=\alpha,\beta=1,2,3,
                                         \end{array}\right.
                                         \nonumber\\[1mm] &&
\calP ^L_{\mu\nu} =\delta_{\mu\nu}-\dfrac{q_\mu q_\nu}{q^2}-\calP ^T_{\mu\nu}.
\end{eqnarray}
The gap equation has  the same form as in case of $T=0$, Eq. (\ref{sde}), except that within the Matsubara formalism
the integration over $k_4$ is replaced by the summation over the corresponding  frequencies, formally
\begin{eqnarray}
\int\dfrac{d^4 p}{(2\pi)^4} \longrightarrow T\sum\limits_{n=-\infty}^{ \infty} \int\dfrac{d^3 p}{(2\pi)^3}.
\end{eqnarray}
Then the system of equations for $A,B$ and $C$ to be solved reads (cf. also Ref. \cite{FischerRenorm})
\begin{eqnarray}
A(\bp^2,\omega_n^2) &&=1+\dfrac{4}{3}T  \sum\limits_{m=-\infty}^{ \infty} \int\dfrac{d\bk}{(2\pi)^3}
\left\{ 2\qk \left( 1-\pk\right)\sigma_A(\bk,\omega_m) D^T(\bq,\Omega_{nm})\right. +
\nonumber \\ &&
\left. \left[\pk\sigma_A(\bk,\omega_m) + 2\dfrac{\Omega_{mn}}{q^2}\left(1-\pk\right)\omega_m \sigma_C(\bk,\omega_m)-
\right. \right .\nonumber
\\ &&
\left.\left.
-2\dfrac{\Omega_{mn}^2}{q^2}\qk\left(1-\pk\right)\sigma_A(\bk,\omega_m)\right] D^L(\bq,\Omega_{nm},m_g)\right\},\label{AA}\\[2mm]
B(\bp^2,\omega_n^2) &&=m_q+\dfrac{4}{3}T  \sum\limits_{m=-\infty}^{ \infty} \int\dfrac{d\bk}{(2\pi)^3}
\left[\!\!\!\!\phantom{\frac11} D^L(\bq,\Omega_{nm},m_g) +2D^T(\bq,\Omega_{nm},0)\right]\sigma_B(\bk,\omega_m),\label{B}
\\[2mm]
C(\bp^2,\omega_n^2) &&=1+\dfrac{4}{3}T  \sum\limits_{m=-\infty}^{ \infty} \int\dfrac{d\bk}{(2\pi)^3}
\left\{ 2\dfrac{\omega_m}{\omega_n}\sigma_C(\bk,\omega_m) D^T(\bq,\Omega_{nm},0)+\right.\nonumber\\[1mm] &&
\left.\left[-\left
( 1-2\dfrac{\Omega_{mn}^2}{q^2}\right )
 \dfrac{\omega_m}{\omega_n}\sigma_C(\bk,\omega_m)+
2\dfrac{{\bf q k}}{q^2}\dfrac{\Omega_{nm}}{\omega_n}\sigma_A \right] D^L(\bq,\Omega_{nm},m_g)\right\},
\label{C}
\end{eqnarray}
where $\bq=\bp-\bk $ and the propagator functions $\sigma_F=\sigma_F(\bk,\omega_m)$ are defined by
\begin{eqnarray}
\sigma_F(\bk,\omega_m)=\dfrac{F(\bk,\omega_m)}{
\bk^2 A^2(\bk,\omega_m) + \omega_m^2 C^2(\bk,\omega_m)+B^2(\bk,\omega_m)}
\end{eqnarray}
for  $F=A,B$ and $C$.
The form of the  interaction kernel  is taken  the same as at $T=0$, i.e.
both the transversal and longitudinal parts consisting of two terms~-~the
 infra-red and ultraviolet ones. The information on these
 kernels is even more sparse than in the case of $T=0$. While the effective parameters
 of the kernel in vacuum can be adjusted to some known experimental data,
 e.g. the meson mass spectrum from the BS equation, at finite temperature one
 can rely  on results of
 QCD calculations, e.g. by using results of the nonperturbative lattice calculations.
 There are some indications, cf.~\cite{tereza},
 that at low temperatures the gluon propagator is insensitive to the temperature impact,
  and the interaction can be
 chosen  as at $T=0$ with $D^T=D^L$ \cite{kitaizyRpberts}. However, in a hot and/or dense
medium  the gluon is also subject to medium effects and thereby becomes effectively massive with
finite transversal (known also as the Meissner mass) and longitudinal (Debye or electric)
 masses. Generally, these masses  appear  as independent parameters with
 contributions depending on the considered  process~\cite{Tagaki1}.  {The
role of the Meissner masses   in the tDS equation
at zero chemical potential  is not yet well established and requires separate
investigations. This is beyond the goal of the present paper where
only the Debye mass, $m_g$, is considered. It should be
noted that, independently of the value of the chemical potential, in most
approaches based on the tDS equation within the rainbow approximation it is also  common practice
to ignore the effects of Meissner masses. This is inspired by the results
 of a tDS equation analysis in the high temperature and density
   region~\cite{meiss} which report that the
 Meissner mass  is of no importance in tDS equation. }  At this level
the Debye mass is the only $T$ depending  part of the kernel. The Debye mass is well defined in the weak-coupling
regime. In~\cite{AlkoferDebye,Tagaki,fischerPRD90,kalinovsky}  it was found that the  $T$-dependence
of the Debye mass is in leading order
\begin{eqnarray}
m_g^2=\alpha_s
\dfrac{\pi}{3}\left[ 2 N_c+N_f\right]T^2,\label{Debye}
\end{eqnarray}
where $N_c$ and $N_f$ denote the number of active color and flavor degrees of freedom, respectively;
  the running coupling  $\alpha_s$ in the one-loop approximation  is
\begin{eqnarray}
\alpha_s(E)\equiv \dfrac{g^2(E)}{4\pi}=f(E) \dfrac{12\pi}{11 N_c-2N_f}
\end{eqnarray}
with $E$ being the energy scale. For the temperature range considered in the present
paper we adopt $f(E)\to 2$, whichoften employed~\cite{AlkoferDebye,Tagaki,fischerPRD90,kalinovsky}
 choice for the Debye mass in the tDS equation. It should be noted, however,  that such a choice
 of $f(E)$ is not unique. It may vary in some interval, in   dependence on
  the employed method of  infra-red regularization~\cite{Tagaki,kapusta}.
  Since the Debye mass enters as an additional energy parameter in
  $q^2=\bq^2+\Omega_{mn}^2+m_g^2$, which determines the Gaussian form of the IR part of
  the interaction~(\ref{phenvf}),
  an increase of   $m_g^2$ results in a shift of the tDS solution towards lower temperatures
   leaving, at the same time, the shape  of the solution practically  unchanged. Accordingly, smaller values of $m_g^2$ shift the solution
  towards larger temperatures. Our numerical calculations show that a decrease of $m_g^2$ by a factor of 2 results
  in a $\sim 15\%$ shift of the solution to larger temperatures.
The transversal and longitudinal parts of the interaction kernel~(\ref{kernel})
can be  cast in the form
\begin{eqnarray} &&
D^T(\bq,\Omega_{mn},0)=D_{IR}(\bq^2+\Omega_{mn}^2)+D_{UV}(\bq^2+\Omega_{mn}^2), \\ &&
D^L(\bq,\Omega_{mn},m_g)=D_{IR}(\bq^2+\Omega_{mn}^2+m_g^2)+D_{UV}(\bq^2+\Omega_{mn}^2+m_g^2).
\end{eqnarray}
In the present paper we use several sets of parameters for the interaction kernel (\ref{phenvf}):\\
1) $\omega=0.5$ GeV , $D=1$ GeV$^2$, $m_u=5$ MeV, $m_s=  115$ MeV;
 results with these parameters are denoted as AWW  (IR term only)~\cite{Alkofer} and MT1 (IR+UV terms)~\cite{MT}.\\
2) $\omega=0.4$ GeV and $D=0.93$ GeV$^2$, $m_u=5$ MeV, $m_s= 115$ MeV;
  MT2~\cite{MT}.\\
3)  AWW, MT1 and MT2 with a modified parameter $D$ making it dependent on temperature;
 at low $T$ it remains constant, equal to the values used in the
 AWW, MT1 and MT2 sets,  while at large temperatures, where the IR contribution
 is expected to be screened, the parameter  $D$ becomes a decreasing function of $T$. In this case,
 since the IR term vanishes, the AWW model is not applicable. It should be noted that
 all these models provide  values for the vacuum quark condensate in a narrow corridor,
 $-\la q\bar q\ra_0 = (0.0145 - 0.0159)$ GeV$^3$, and
 the correct $\pi$ and $\rho$ meson masses as quark-antiquark bound states~\cite{Blank:2011ha}.

  \section{Solutions of the \lowercase{t}DS equation in the chiral limit}
\label{sec3}
\subsection{Order parameters}\label{order}
As seen from Eq. (\ref{B}) in the chiral limit, i.e. at $m_q=0$,
the  trivial solution $B(\bp,\omega_n)= 0$ is possible,
known as the Wigner-Weyl mode.  It is of  separate interest since this   is the
case where the dynamical chiral symmetry breaking is completely disabled.
In the present paper, however, we are interested in solutions
with finite dynamical quark masses given by the ratio  $B/A$, which enter the
 BS equation and determine accordingly
the  hadron  bound states in a heat bath.
Therefore,  we will not consider the Wigner-Weyl mode solution and focus instead on
$B\neq 0$ (Nambu-Goldstone mode). It should  be also noted that even for $B\neq 0$  the sign of  $B$
is not defined.  Equation~(\ref{B}) is invariant under  $B\to -B$.  The sign of the
solution can be fixed only by fixing the sign of  the initial conditions for $B$.
Here  we consider  positive values of $B$.

{We solve numerically the system of equation (\ref{AA})-(\ref{C}) by an iteration
procedure. Since the UV term in the MT1 and MT2 models is logarithmically divergent, a regularization
of the integral over the internal momentum and summation over $\omega_n$ is required.
Usually, at $T=0$ one employs an $O(4)$ invariant cutoff $\Lambda$. The dependence of
the solution on $\Lambda$ is removed by choosing a subtraction scheme defined at a
renormalization point $\mu\le \Lambda$ such that
$A(p^2=\mu^2,\Lambda^2)=1; \ B(p^2=\mu^2,\Lambda^2)=m$. In an analogous way one performs the renormalization procedure
at finite temperature  $T$~\cite{FischerRenorm,MTrenorm}. The only difference is that the internal momentum $\bf k$
is restricted  by the condition ${\bf k}^2 + \omega_n^2 \le \Lambda^2$.
 At each iteration step this requires an interpolation of the previous solution to  define
the new Gaussian mesh for ${\bf k}_{max} = \sqrt{\Lambda^2 - \omega_n^2}$.
In our calculations we employ  a cubic spline interpolation procedure  and a mapping
\begin{eqnarray}
          k=k_0\frac{1 + x}{1-x}
\end{eqnarray}
for the Gaussian integration with  $k_0=0.85$ for a mesh of 64 nodes. This provides a rather large cutoff
$\Lambda={\bf k}_{max}\simeq 10^{3} $ GeV/c.
}
The summation over $\omega_n$ is truncated  at a
large value of $n=N_{max}$, where in our calculations $N_{max}\sim 320$ for
 low temperatures and $N_{max}\sim 250$ at temperatures $T > 80 -100 $ MeV are utilized.
 In Figs.~\ref{fig1} and ~\ref{fig2} we exhibit the solution of the system (\ref{AA})-(\ref{C}) for the
 lowest Matsubara frequency $\omega_0=\pi T$ in dependence on the momentum $|\bp|$ at
 low temperature ($T=5 $ MeV, Fig.~\ref{fig1})
 and  higher temperature ($T=100 $ MeV, Fig.~\ref{fig2}).  A comparison with the
 vacuum solution~\cite{our2013}  shows
 that qualitatively there is no difference of  the solutions  at finite $T$. To emphasize the dependence
 on the effective parameters $D$ and $\omega$, in Figs.~\ref{fig1} and \ref{fig2} we present results of calculations
 for the two different sets. { In Fig. 1, left panel, the solutions
 $A({\bf p},\omega_0)$ are represented by
 solid and dashed curves,  while the solutions $C({\bf p},\omega_0)$  by
dotted and dash-dotted curves  for MT1 and MT2 sets, respectively. The solutions $B({\bf p},\omega_0)$, left panel,
 are for MT1 (solid curve) and NT2 (dashed curve).} The dependence on the values of
  the effective parameters is seen only at low momenta. In this region, the dependence
  is mainly determined  by the slope parameter $\omega$, cf.  Eq.~(\ref{phenvf}),
 which is quite different for the two employed  parameter sets. At larger momenta (cf. Fig.~\ref{fig2}),  the common asymptotics
 is approached  already at $|\bp| > 10$  GeV/c.
\begin{figure}[!ht]
\includegraphics[scale=0.5 ,angle=0]{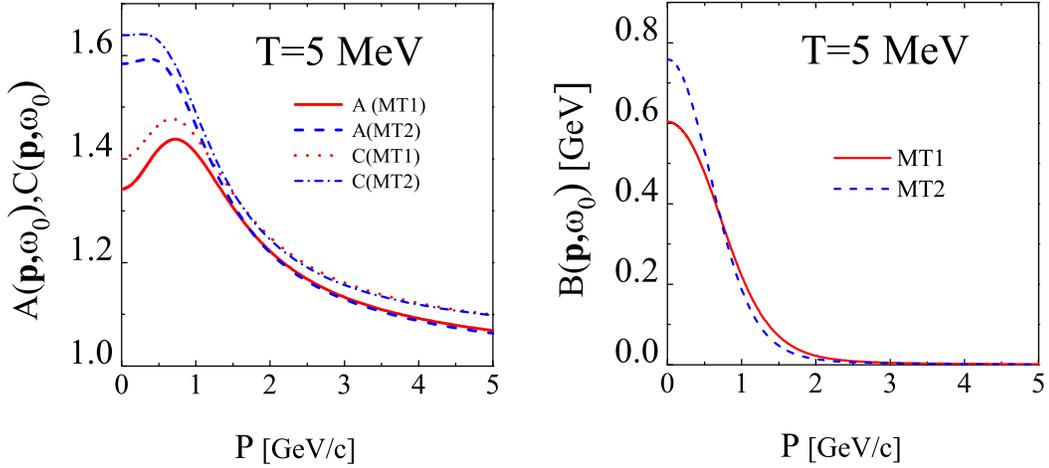}
\caption{(color online)
Solutions of Eqs.  (\ref{AA})-(\ref{C}) for the lowest
Matsubara frequency $\omega_0$ at $T=5$ MeV in the chiral limit, $m_q=0$.
The solution for  $\omega=0.5$ GeV and $D=1$ GeV$^2$  is labeled as MT1, while
for the parameters $\omega=0.4$ GeV and $D=0.93 \ GeV^2$ as  MT2.
Both sets of parameters include the IR and UV terms.
 In the left   panel  the solution for
$A(\bp,\omega_0)$ is depicted by solid (MT1) and dashed
(MT2) curves, and $C(\bp,\omega_0)$ by dotted and dash-dotted curves,
respectively, while
the right panel exhibits the function $B$ for MT1 (solid)  and MT2 (dashed)
 kernels.
}
\label{fig1}
\end{figure}
 \begin{figure}[!ht]
\includegraphics[scale=0.5 ,angle=0]{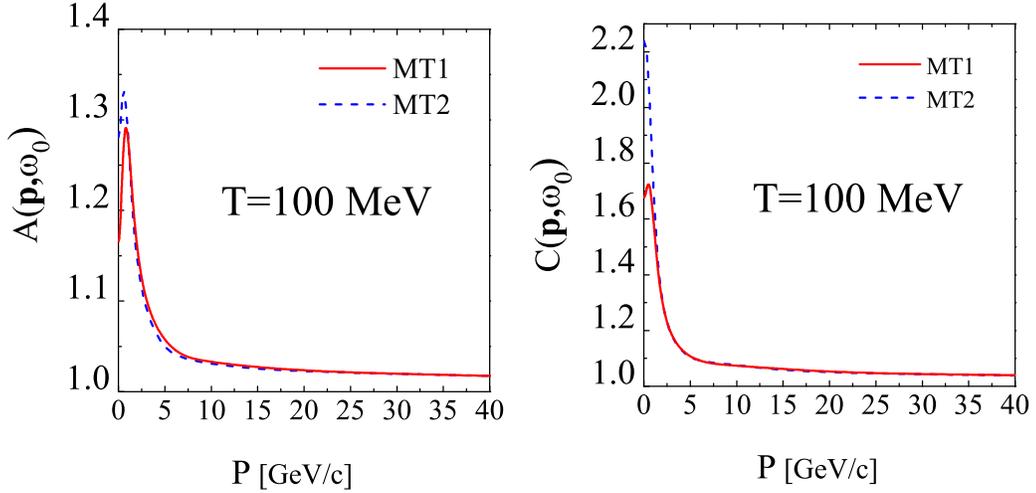}
\caption{(color online)
Solutions of Eqs. (\ref{AA})-(\ref{C}) in the chiral limit,
$m_q=0$,  for the lowest Matsubara frequency
 $\omega_0=\pi T$ at $T=100$ MeV  for $A(\bp,\omega_0)$ (left panel) and $C(\bp,\omega_0)$ (right panel).
Solid (dashed) curves are for MT1 (MT2).
Since the quantity $B$ becomes negligibly small, $B< 10^{-10}$, at $T\ge \ 100$ MeV it
is not exhibited in this figure.}
\label{fig2}
\end{figure}
 The dependence of the solution on  the temperature is of particular interest. It is known that in dense and hot matter
there may occur different kind of phase transitions.
In SU(3) gauge theory, the deconfinement transition is of first order at
$T_c = {\cal O}(270)$ MeV related to the center symmetry, while in 2+1 flavor QCD with
physical quark masses  it is a cross-over at
$T_c={\cal O}(150)$ MeV, see~\cite{bazavov,Fodor_glue,WuBp_quarks}. At non-zero baryon density, the liquid-gas phase
transition at $T_c={\cal O}(20)$ MeV matters, and a critical end point of an additional
 first-order phase transition curve
is still hypothetical. According to the Columbia  plot~\cite{Columbia}, two-flavor QCD in
the chiral limit displays a first-order deconfinement and chiral restoration transition.
The corresponding quantity which characterises such transitions
is known as order parameter of the considered media.
Natural candidates to be considered as
order parameters or elements thereof
are the mass function $B(\bp,\omega_n)$ and the quark condensate $\la q\bar q \ra $, being  an integral
characteristics of the mass $B( \bp,\omega_n)$ too. Order parameters determine the so-called critical
temperature $T_c$  or the cross-over region which will serve as an  indicator for a possible (phase) transition.

At high enough temperatures one expects a chiral restoration. This means that at a certain
high value of the temperature the mass function $B$ should vanish, indicating
a   possible phase transition in the hot matter. The lowest
 temperature at which $B=0$ holds is called   the critical temperature $T_c$, i.e.
for the mass function $B(\bp,\omega_n)$ the critical temperature can be defined as that value of $T$ at which the
solution $B(\bp,\omega_n)$ vanishes. Analogously, for the chiral condensate, $T_c$ can be determined also
as the temperature above which $\la q\bar q \ra $ vanishes. In principle, these two critical temperatures can be slightly different.

The chiral condensate is defined by
 \begin{eqnarray}
 \la q \bar q \ra &&=-4N_c T\sum\limits_{n=-\infty}^\infty
 \int\dfrac{d\bp}{(2\pi)^3} Tr\left[
 S(\bp,\omega_n)\right ]\nonumber\\ && =
 -4N_c T\sum\limits_{n=-\infty}^\infty\int\dfrac{d\bp}{(2\pi)^3}
 \dfrac{B(\bp,\omega_n)}{\bp^2 A^2(\bp,\omega_n)+\omega^2_n C^2(\bp,\omega_n)+B^2(\bp,\omega_n)},
 \label{condensate}
 \end{eqnarray}
 where the trace is performed in spinor space.
 In Fig.~\ref{fig3} we present results of calculations of the $T$-dependence of the mass solution $B$
  (left panel), and the normalized chiral condensate $\la q \bar q \ra $ (right panel) in the chiral limit  $m_q=0$.
 \begin{figure}[!ht]
\includegraphics[scale=0.5 ,angle=0]{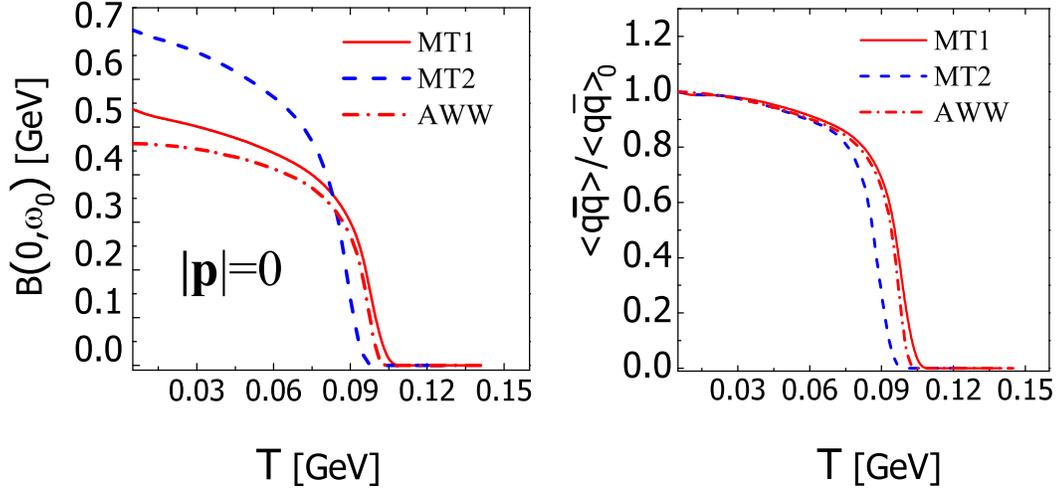}
\caption{(color online)
Solutions  $B(\bp,\omega_0 )$ from (\ref{B})
in the chiral limit, $m_q=0$,  for the lowest Matsubara frequency  (left panel) and
the chiral condensate (\ref{condensate}) normalized at low $T$
 (right panel) as  functions of $T$.
 The dashed and solid curves are for the same  effective
 parameters as in  Fig.~\ref{fig1}.}
\label{fig3}
\end{figure}

\begin{figure}[!ht]
\includegraphics[scale=0.5 ,angle=0]{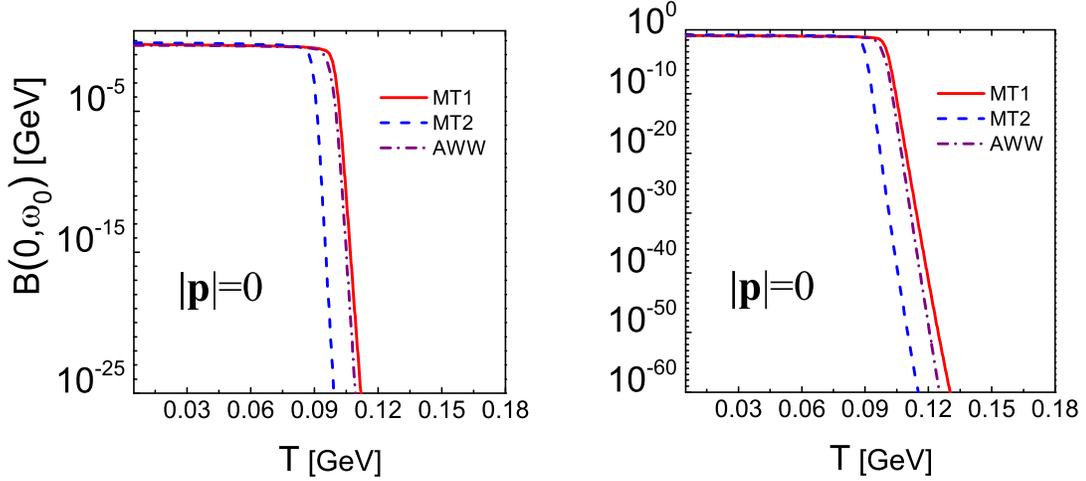}
\caption{(color online)
Solution  $B(\bp,\omega_0 )$  from Eq. (\ref{B}), depicted in a log-scale,
in the chiral limit, $m_q=0$,  for the lowest Matsubara frequency.
 The   solid, dashed and dot-dashed curves are for  the MT1, MT2 and AWW models,
 respectively. The  left panel  exhibits the sharp dropping of
  $B(\bp,\omega_0 )$ in the vicinity of
 $T \sim 100-105$ MeV which continues when extending the scale
 to even
 smaller values, as illustrated in
the right panel. From the right panel one would infer that
  $T_c \approx 110$ MeV for the MT2 model and $T_c \approx 130$ MeV for
  the MT1 model.}
\label{fig3a}
\end{figure}

 One infers from this figure that  in a
large range of $T$ the solution $B(0,\omega_0)$ is a smoothly decreasing  function of $T$,
except for  a narrow interval where  $B(0,\omega_0)$ sharply decreases towards  zero, as  seen in  Fig.~\ref{fig3a} where,
for the sake of better determination of $T$ for which   $B(0,\omega_0)\to  0$, the solution
  is presented in a log-scale.  It should be stressed  that for
very   small values of $B$ ($B\le 10^{-20}$) the convergence of the iteration procedure for  Eq. (\ref{B}) becomes rather poor.
 One needs to   increase essentially  the number of iterations in order to achieve the same
 accuracy in the whole range of considered temperatures. In addition, the actual accurate
 calculations in the neighbourhood of $T_c$
 are restricted by   numerical manipulations with quantities close to the machine zero.
 These numerical effects hinder a precise determination of $T_c$ in the chiral limit, $m=0$.
 In our calculations, we analyse the values of $B$ below $ 10^{-60}$ to
 determine  the interval for $T_c$. An inspection of the numerical results shows that, for the MT1
model, the critical temperature is about 130\ MeV, while for the MT2 model one has about 110 MeV. An analogous
determination of $T_c$ from the chiral condensate provides  slightly different values, e.g.
$T_c=128$ MeV for  MT1  and $T_c=105$ MeV for the MT2 model. The AWW set provides values
close to the MT1 model. The obtained values are by $ 20\% $  smaller than that
obtained from QCD calculations~\cite{yoki}, which report  a cross-over temperature $T_c$ in
the range $T_c\sim [145\ldots 170 ] $
MeV which is now narrowed to ($154\pm 9$)    MeV~\cite{ding},
 however, for 2+1 flavor QCD with physical masses.}
It should be also noted that the general feature of the quark condensate, as a function of
the temperature below the chiral transition limit,
is established  in a model independent way~\cite{hatsuda,gerber} by the low-temperature expansion
$
\la q\bar q\ra=\la q\bar q\ra_0 [1-T^2/8f_\pi^2-{\cal O}(T^4)],
$
where $f_\pi$ is the pion decay constant in the chiral limit ($f_\pi\approx 93$ MeV). Our results in
Fig.~\ref{fig3} are in a qualitative agreement with that. It should be emphasized that the above
quoted values for $T_c$ stem from the inspection of the numerical results of $B$ and $\la q\bar q\ra$ at
small values. On the linear scale in Fig.~\ref{fig3} however, one infer instead values of about $120$ MeV or
even less.
    \subsection{Spectral representation above ${\bf T}_c$}\label{spectral}
Another important quantity characterizing the hot matter is the spectral representation of the
retarded quark propagator.
  The Euclidean   propagator   can be transferred  to  Minkowski space by an analytical continuation
  of the solution of the gap equation  to real energies,
  \begin{eqnarray}
  S^M(\bp,\omega)=\left. S(\bp, i\omega_n)\right |_{i\omega_n\to \omega+i\eta}.
\end{eqnarray}
In  Minkowski space, the dispersion relation for the quark propagator
determines  the spectral representation $\rho(\bp,\omega)$ which is directly related
to  the imaginary part of the propagator, $\rho(\bp,\omega)=-2\Im   S^M(\bp,\omega)$.
It means that in Euclidean space the same spectral density $\rho(\bp,\omega)$
can be associated to the retarded quark propagator
\begin{eqnarray}
S(\bp,i\omega_n)=\frac{1}{2\pi}\int\dfrac{\rho(\bp,\omega')}{i\omega_n-\omega'}d\omega'.
\label{rhon}
\end{eqnarray}
From this   the importance of studying $\rho$ can be inferred. Note that, since the
spectral density characterises the propagation of a particle, the dispersion relations in our case
are meaningful only in the  (deconfinement) region $T>T_c$,
where  quarks can be treated  to some extent as quasi-particles.
Recall that in Minkowski space the propagator of a free particle  can be written as
\begin{eqnarray}
S(p)=S_+ + S_-=\frac{\Lambda_+}{\omega-E_\bp}\gamma_0 + \frac{\Lambda_-}{\omega+E_\bp}\gamma_0,
\end{eqnarray}
where $E_\bp=\sqrt{\bp^2+m^2}$ is the energy of the particle and
$\Lambda_\pm= {E_\bp \pm \gamma_0 [\bgam \bp +m]}/{2E_\bp}$ are the
projection operators on positive and negative energy solutions, respectively,
obeying
$\Lambda_\pm\Lambda_\pm = \Lambda_\pm, \ \Lambda_++\Lambda_-=1,
\ \Lambda_+\Lambda_-=0$.    Therefore, for a free quark
\begin{eqnarray}
\rho_\pm (\omega)=\frac12 \delta(\omega\mp E\bp),
\label{freerho}
\end{eqnarray}
i.e. the spectral functions $\rho_\pm (\omega)$ characterize the propagation of
quasi-particles with positive (normal) and negative (abnormal) energies.
 This can serve as a hint in parametrizing the spectral density $\rho(\bp,\omega)$ in
 Euclidean space.
Owing to parity and rotation  symmetries, the Dirac
structure of the quark spectral function at finite temperature
is in general decomposed as
\begin{eqnarray}
\rho(\bp,\omega)=\rho_4(\bp,\omega)\gamma_4+\rho_v(\bp,\omega)(\bgam \bp)+\rho_s(\bp,\omega).
\end{eqnarray}
In the present paper we focus to two particular  cases. \\
 (i) Chiral limit, where the scalar, or "mass",  part $\rho_s(\bp,\omega)$ vanishes.
 In this case, the spectral function at negative energies describes  the so-called plasmino
 mode~\cite{kitazavaPRD80}.
 In the chiral limit the projection operators $\Lambda_{\pm}$ are of a particularly
 simple form and the spectral density can be written as
 \begin{eqnarray}
 \rho(\bp,\omega)= \rho_+(\omega)\frac{(1+i\gamma_4\bgam\hat\bp)}{2} +
 \rho_-(\omega)\frac{(1-i\gamma_4\bgam\hat\bp))}{2},
\label{decRho}
\end{eqnarray}
where $\hat\bp\equiv \bp/|\bp|$.
 \\
 \noindent
 (ii) Zero momenta: The projection operators are
 $\Lambda_{\pm}=\dfrac{1\pm \gamma_4}{2}$ and
 \begin{eqnarray}
 \rho(\omega,\bp=0)= \rho_+(\omega)\frac{1+\gamma_4}{2} +\rho_-(\omega)\frac{1-\gamma_4}{2}.
 \label{decompRho}
\end{eqnarray}

Note that at zero momenta the energy $E$ of the quark  can be associated to a mass $m_T$ which
in literature is referred to as  the thermal mass, a subject of many investigations within
lattice QCD, cf.~\cite{kitazavaPRD80}.
Results of such calculations are often considered as "experimental" data for the corresponding quantity.
This is an important issue, since the model calculations of $m_T$ can be related to "experimental"
data and to serve  as a guide in fixing the  phenomenological parameters and to estimate the
applicability of the model  which is based directly on  parametrisations and scale settings by vacuum
meson physics.
It can be shown that the quark propagator can be written in the same form (\ref{decRho}).
So, in the chiral limit one can note
 \begin{eqnarray}
 S(\bp,\omega_n )= S_+(\bp,\omega_n)\frac{(1+i\gamma_4\bgam\hat\bp)}{2} {\gamma_4} +
 S_-(\bp,\omega_n)\frac{(1-i\gamma_4\bgam\hat\bp))}{2} {\gamma_4},
\end{eqnarray}
where
\begin{eqnarray}
S_{\pm}(\bp,\omega_n)=-\dfrac{i\omega_n C(\bp,\omega_n)\pm |\bp| A(\bp,\omega_n))}
{\omega_n^2 C^2(\bp,\omega_n) +\bp^2 A^2(\bp,\omega_n) }.
\label{Spm}
\end{eqnarray}
If one writes the dispersion relations for the model propagators (\ref{Spm})
 \begin{eqnarray}
S_{\pm}(\bp,\omega_n))=\frac{1}{2\pi}\int \dfrac{\rho(\bp,\omega)}
{ i\omega_n-\omega} d\omega,
\label{dr}
\end{eqnarray}
 then by inverting (\ref{dr}) one can obtain the (model) spectral density  $\rho(\bp,\omega)$.
Note that in model calculations the problem of inverting expressions like (\ref{dr}) is ill posed.
Nevertheless,  instead of inverting the equation (\ref{dr}),
one can suggest a reliable  parametrization for  $\rho(\bp,\omega)$ which allows an analytical calculation
of the integral over $\omega$ and then
to  minimize the quantity
\begin{eqnarray}
\Delta_N=
\frac{1}{2N+1}\sum\limits_{n=-N}^N
\left|S_{\pm}(\bp,\omega) - \frac{1}{2\pi}\int\dfrac{\rho_\pm(\bp,\omega)}{i\omega_n-\omega}d\omega\right |^2,
\label{minim}
\end{eqnarray}
where the integral   in Eq.~(\ref{minim}) must be  preliminarily carried out analytically  to
leave the dependence only on $\omega_n$ and effective parameters.
In such a way one can find the effective parameters and estimate the behaviour of $\rho(\bp,\omega)$.

The simplest parametrization for the spectral function at
finite $T$ is suggested by the case of a free quark propagator~(\ref{freerho}), i.e.
one can expect that $\rho(\bp,\omega)$ exhibits two maxima. For the two-pole parametrization
the spectral function reads
\begin{eqnarray}
\rho_+(\bp,\omega)=Z_1(|\bp|)\delta({\omega-E_1} )+ Z_2(|\bp|)\delta(\omega+E_2).
\label{twopoles}
\end{eqnarray}
With such a parametrization the spectral function $\rho(\omega,\bp)$
describes the propagation of quasi-particles with the
 positive energy $E_1$
and aniti-particles with negative energy $E_2$;
the second term in~(\ref{twopoles}) is known also as the plasmino mode~\cite{kitazavaPRD80}.
The weights $Z_{1,2}(|\bp|)$ of the normal and plasmino modes play an  important role in
estimating the phase transitions in hot matter.
At zero momenta, one has $E_1=E_2$ and $Z_{1}(|\bp|=0) =Z_2(|\bp|=0) = 1/2$.
In this case the energy parameters determine the thermal masses, $E_{1,2}=m_T$, which
 are predicted~\cite{kitazavaPRD80} in lattice QCD to be an increasing function of $T$ and at
$T/T_c \ge 2$ to be  $m_T\sim 0.9 \ T$. This  important "experimental" result   may be used
in choosing the model kernels of the  tDS equation.
As the momentum $|\bp|$ increases one expects that the plasmino mode vanishes and $Z_{1}(|\bp|)\to 1$.

In our  calculations we use the Levenberg-Marquardt algorithm for minimization of $\Delta_N$ in Eq. (\ref{minim}).
Results of calculations are exhibited in Fig.~\ref{fig4}.
\begin{figure}[!ht]
\includegraphics[scale=0.5 ,angle=0]{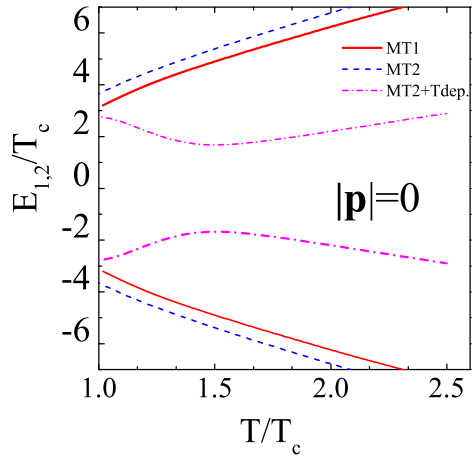}
\caption{(color online)
The scaled thermal masses as function of $T/T_c$ in the chiral limit.
 The dashed and solid curves are obtained with two different sets of the effective
 parameters, MT1 and MT2. The dot-dashed curve is obtained by the modified,
  $T$-dependent kernel of Eq. (\ref{DotT}).}
\label{fig4}
\end{figure}
It can be seen that both, MT1 and MT2 models
(solid and dashed curves in  Fig.~\ref{fig4}), provide increasing functions
of $T$. However, the absolute values at $T\ge 2 T_c$ are far from that predicted by
QCD~\cite{yoki}. It implies that, while at low temperatures
the two models with vacuum parameters, maintain a satisfactory description of the quark propagators
 at  $T>T_c$,
the interaction kernel requires  modifications, cf. Ref.~\cite{robLast,kitaizyRpberts}.
Such modifications are inspired by the fact that, at  sufficiently large temperatures, thermodynamics
should be describable in terms of a weakly interacting
quark-gluon gas, and at asymptotically large  temperatures all thermodynamic
quantities should converge to the ideal gas limit (for  a discussion of
 the lattice QCD approaching to the perturbative limit see, e.g.  Ref.~\cite{bazavov}).
It means that, at large temperatures,  the
 IR term  in the interaction kernel  must diminish or even  vanish.
A first attempt to modify the interaction kernel was done in Ref.~\cite{kitaizyRpberts}, where
the weight $D$ of the IR term above  $T_c$ is abruptly  (via a step function) replaced
by a phenomenological, $T$-dependent decreasing function.
In the present paper we suggest another  modification of the IR term which smoothly
decreases at large $T$ and, at the same time,  does not affect the IR term at low $T$.
To do so, we introduce a suppression function $f_1(T)$ which has a  Heavyside step-like
behavior at temperatures $T\sim T_c$,
\begin{eqnarray}
D\to D(T)=D\ f_1(T) =D\ \frac12 \left[ 1+\tanh\left (-\frac{ T-T_p }{\beta}\right )\right],
\label{DotT}
\end{eqnarray}
where the additional adjustable parameters are $T_p\sim T_c$ and $\beta$ as some
diffuseness of the IR interaction.
With such a parametrization,  at $T \ll T_p$ the weight $D$ of the IR
 part is as in the vacuum,  $D(T)=D$, and at $T\gg T_c$, $D(T)\to 0$.
In  our calculations
we adopt $T_p=130-140$ MeV and $\beta=30$ MeV, which assures for both, MT1 and MT2 models,
 a reasonable behaviour of the thermal masses
$m_T\sim T$ at $T\ge  2T_c$, see Fig.~\ref{fig4} (cf. also Ref.~\cite{yoki}).

Another important characteristic is the behaviour of the plasmino mode as a
function of the momentum $\bp$.
We find, cf. also~\cite{kitaizyRpberts}, that, while the energy of the normal mode monotonously  increases
with $|\bp |$ (as it should be), the  plasmino mode decreases up to a minimum value,
than sharply increases approaching the normal mode at large $|\bp |$, see Fig.~\ref{fig5}-left panel.
This does not mean at all that the role of plasmino mode is increasing too and becomes of the same
importance as the  normal one. Instead, the weight $Z_2(|\bp |)$ sharply decreases with increasing $|\bp |$,
vanishing at large $|\bp |$, cf. Fig.~\ref{fig5}-right panel. The local minimum of the plasmino
mode is related to the Van Hove singularity.
\begin{figure}[!ht]
\includegraphics[scale=0.5 ,angle=0]{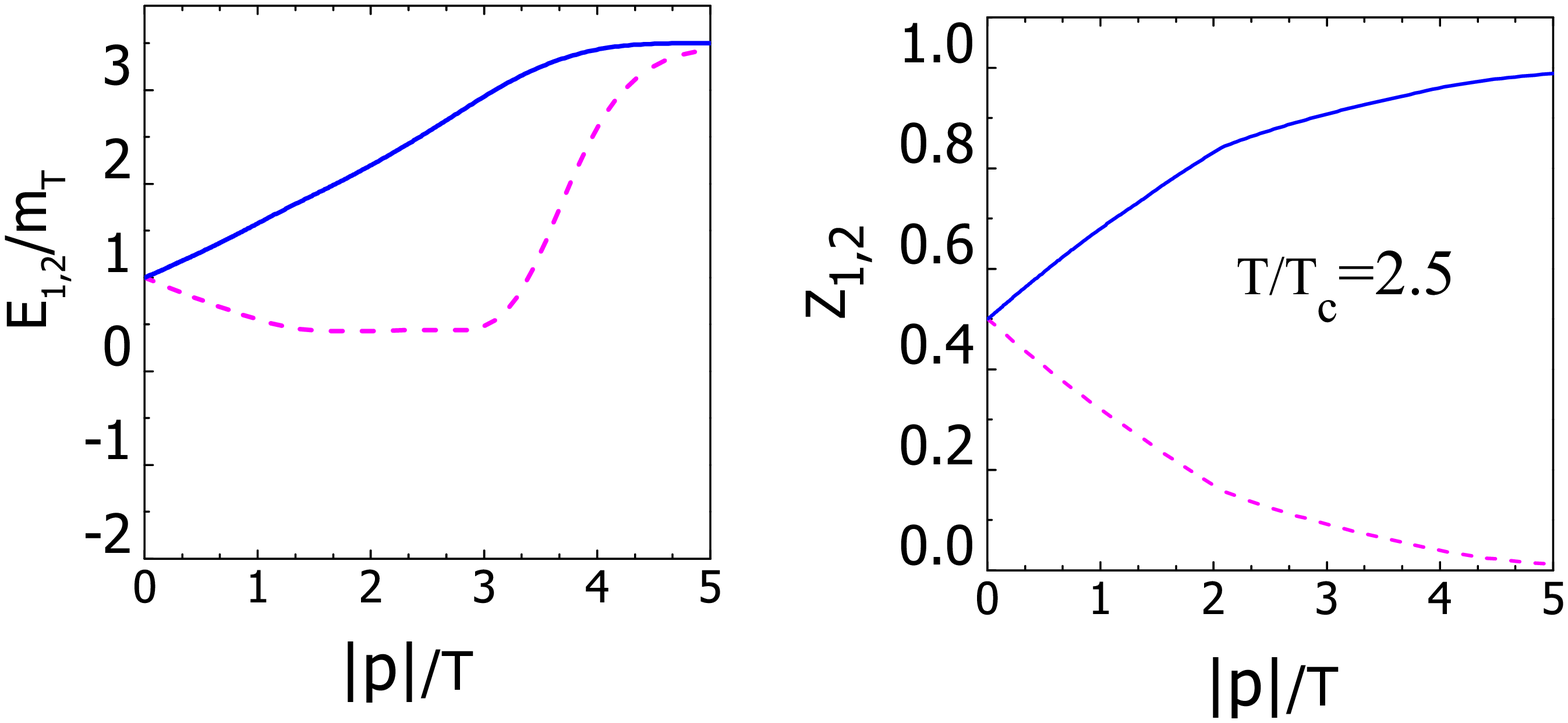}
\caption{(color online)
Left panel: Energy of the normal (solid curve) and plasmino mode (dashed curve) as a function
of the scaled momentum computed with
the $T$-dependent interaction~(\ref{DotT}). Right panel: The corresponding weights of
the normal (solid curve), and plasmino modes (dashed curve) at $T/T_c=2.5$.}
\label{fig5}
\end{figure}

\section{Solution of the \lowercase{t}DS equation at finite bare masses}\label{sec4}
 At finite quark masses the solution of the tDS equation differs from the chiral limit in at least two
 aspects. First, the Wigner-Weyl mode is not longer a solution. Second,
 the integrals over $\bp$ are logarithmically divergent in the UV region. This
 means  that a regularization and renormalization procedure is required.
  To make the results finite  one usually  uses
 the Feynman method by introducing a cut-off parameter $\Lambda$ for the integrals, followed by a subsequent reliable
 subtraction procedure~\cite{FischerRenorm,Fischerrenorm1,rob-1}.
 In approximate  models   after  performing the necessary regularizations and
  renormalizations have been performed, the effective  phenomenological parameters are fixed in such a way
 that  a bulk of  the effects  is already included.
 Numerically it implies that by choosing a large enough cutoff $\Lambda$
 there is no need for further normalizations to solve the tDS equation. In our calculations,
 the integral over $|\bp|$ is performed up to $|\bp |_{max}\sim 10^3$ GeV/c which assures an
 asymptotic behaviour of the solution
 $A(\bp,\omega_n)\to 1$, $B(\bp,\omega_n)\to m_q$ and $C(\bp,\omega_n)\to 1$  with an accuracy better than $\sim 0.1\%$.
  However, in further calculations involving  $A,B,C$ one shall bear in mind that additional divergences
  may become apparent for another kind of calculations, and  regularization procedures may   be still required, as e.g.
  in calculations of the chiral condensate.
At finite quark masses the chiral condensate is in fact quadratically divergent,
cf. Eq.~(\ref{condensate}). This  is
manifestly  seen if one considers the quark condensate at  $T=0$ but $m_q\ne 0$,
\begin{eqnarray}
\la q\bar q\ra_0 =-\int\dfrac{d^4 p}{(2\pi)^4} \sigma_B(p_4,\bp))=
-\frac{1}{8\pi^2}\int
\frac{ \tilde p^3 \ B(\tilde p^2) }{\tilde p^2 A^2(\tilde p^2)+B^2(\tilde p^2)} d\tilde p,
\label{divergent}
\end{eqnarray}
where $\tilde p^2=\bp^2 +p_4^2$. At large values of the momenta
the asymptotic solution of the tDS equation becomes  $A(\tilde p^2)\to 1$, $B(\tilde p^2)\to m_q$
and the  integral (\ref{divergent}) is quadratically divergent. On can regularize it by
subtracting at large momenta the asymptotic quark mass $m_q$.
 Denote $B=\tilde B +m_q$, where at large enough momenta $\tilde p_{max}$ the quantity $\tilde B(p_{max})$ goes to zero
 which implies that at $\tilde  p> \tilde p_{max}$  the dynamical chiral symmetry breaking vanishes,
 i.e. the mass function receives its asymptotic value $B( \tilde p_{max}^2) \simeq m_q$. Now, if the cut off
parameter is chosen  large enough, $\Lambda > \tilde  p_{max}$, then
\begin{eqnarray}&&
\la q\bar q\ra_0 (\Lambda) =
-\frac{1}{8\pi^2} \int\limits_0^\Lambda \tilde p^3 d\tilde p \left[
\frac{  \tilde B(\tilde p^2)+m_q }{\tilde p^2 A^2(\tilde p^2)+B^2(\tilde p^2)} \right]\simeq
\la q\bar q\ra_{ren.} - \frac{1}{8\pi^2} \int\limits_0^\Lambda
\frac{ \tilde p^3 \ m_q  }{\tilde p^2 +m_q^2} d\tilde p\nonumber\\[1mm] && =
\la q\bar q\ra_{ren.} -\frac{m_q}{8\pi^2} \Lambda^2 \int\limits_0^1  \frac{x^3}{x^2+\varepsilon^2}dx
=  \la q\bar q\ra_{ren.}+\frac{m_q}{16\pi^2} \Lambda^2\left .\phantom{\!\!\!\!\!\!\frac11}\left [
x^2-\varepsilon^2 \ln(x^2+\varepsilon^2)
\right]\right |_0^{1}  \nonumber \\ &&
\simeq \la q\bar q\ra_{ren.} -   \frac{m_q}{16\pi^2} \Lambda^2 ,\label{regular}
\end{eqnarray}
where
$\varepsilon^2 = \dfrac{m_q^2}{\Lambda^2}\sim 0$, and the  regularized condensate
$\la q\bar q\ra_{ren.}$ does not depend on $\Lambda$. In obtaining (\ref{regular})
we put $A\sim 1$ and $B\sim m_q$ in the second integral.
Equation (\ref{regular}) illustrates the quadratic divergence of the integral and,
at the same time, hints to how  the subtraction procedure is to be chosen  to eliminate this divergence.
With this in mind, one can  define a renormalized (subtracted) quark condensate as
\begin{eqnarray}
\la q\bar q\ra_0^l-\frac{m_l}{m_h} \la q\bar q\ra_0^h = \la q\bar q\ra_{ren.}^l-\frac{m_l}{m_h} \la q\bar q\ra_{ren.}^h,
\label{renormCondensate}
\end{eqnarray}
 where $m_l$ and $m_h$ denote the
  mass of  light, e.g. $u$, and heavy, e.g. $s$, quarks, respectively.
  At $\dfrac{m_l}{m_h}\ll 1$, equation (\ref{renormCondensate})
  determines the required renormalized, cut-off independent light-quark condensate.
  Exactly the same procedure is applied to determine
  the quark condensate at finite $T$, see also Ref.~\cite{fischerPRD90}.
  The remaining multiplicative divergences can be removed by normalizing to
  quark condensate  at zero temperature.

In Fig.~\ref{fig6} we present the dependence of the
 mass function $B({\bf 0},\omega_0)$ and the chiral condensate $\la q\bar q\ra$
  on the temperature at finite bare masses,
 $m_l=5$  MeV and $m_h=115$ MeV. From the figure one infers that,
 since  $B({\bf 0},\omega_0)$ and the chiral condensate $\la q\bar q\ra $ do not exactly vanish
 at large $T$,  their asymptotic values can not serve for
 a clear-cut   definition of the critical temperature. Instead, one can use
 the method of the maximum of the chiral susceptibility, i.e. the maxima
of the derivatives of  $B$ and/or  $\la q\bar q\ra$ with respect to the quark bar mass, as well as
the inflection point of the mass function or of the condensate, i.e. the maxima of the
corresponding derivatives with respect to the temperature~\cite{fischerPRD90}:
\begin{eqnarray}
\chi_B(T)=\dfrac{d^2 B(0,\omega_0)}{dT^2}; \ \ \chi_{qq}(T)=\dfrac{d^2 \la q\bar q\ra }{dT^2}.
\label{deflection}
\end{eqnarray}
The (pseudo-) critical temperature
$ T_c$ is fixed by the condition  $ \left. \chi_B(T)\right |_{T=T_c}=0$ and/or $\left .\chi_{qq}(T)\right |_{T=T_c} =0$.
\begin{figure}[!ht]
\includegraphics[scale=0.5 ,angle=0]{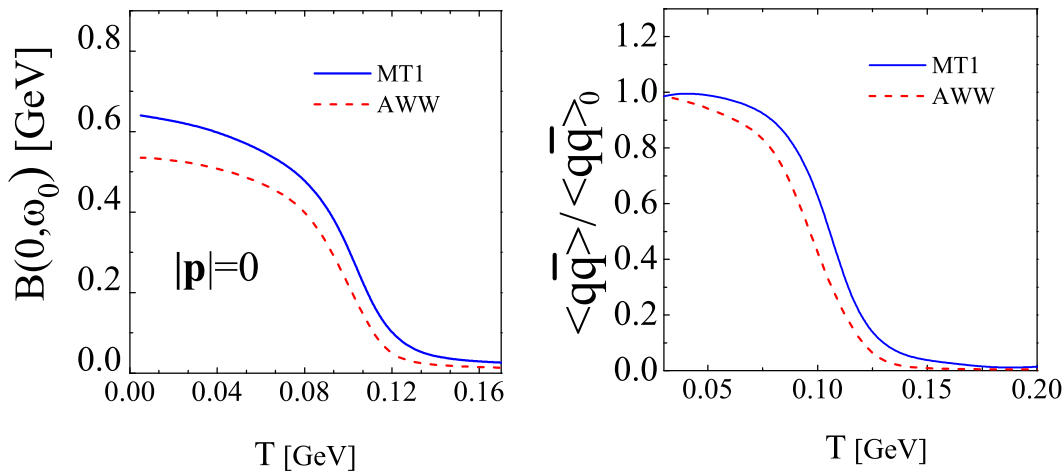}
\caption{(color online)
The solutions $B(\bp=0,\omega_0)$ of the tDS equation
 for the light-quark mass $m_l=5$  MeV   for the lowest Matsubara frequency  (left panel) and  quark
  condensate~(\ref{condensate}) (right panel) as  a function of $T$.
For MT1 (solid)  and AWW (dashed) interaction kernels.
}
\label{fig6}
\end{figure}

\begin{figure}[!ht]
\includegraphics[scale=0.55 ,angle=0]{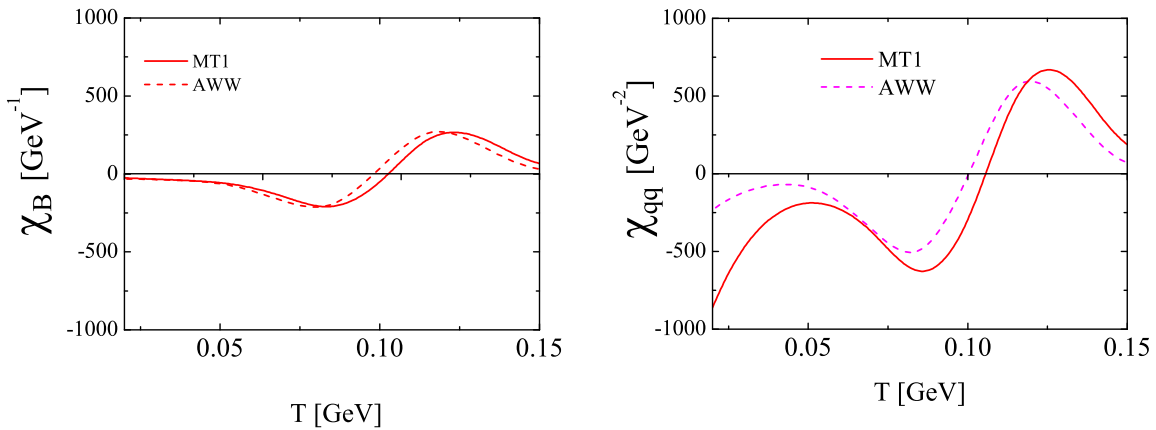}
\caption{(color online)
The inflection points (second derivative with respect to temperature)
for the mass function $B(\bp =0,\omega_0)$ (left panel) and for the normalized quark condensate,
Eq.~(\ref{renormCondensate}) (right panel), as exhibited in Fig.~\ref{fig6}. }
\label{fig7}
\end{figure}

   Figures~\ref{fig6} and~\ref{fig7} clearly demonstrate that the inflection points at finite quark bare masses
provide much smaller (pseudo-) critical temperatures $T_c$;  for all $T$-independent interactions one has
$T_c\sim 100$ MeV, cf. also Ref.~\cite{BlankKrass}.
With the modified interaction, Eq. (\ref{DotT}), for  which the IR term is screened
at large temperatures, $T > 200$ MeV, the positions of the inflection points, which occur
below $T_c={\cal O}(140)$ MeV, remain the same. This implies, for
 a better agreement with the lattice QCD~\cite{yoki,FischerRenorm},
the interaction kernel, for  finite bare quark masses, must acquire an appropriate  dependence on
the temperature also below $T_c$.  { Another important issue of our analysis of the $T$-dependence
of the IR term is that, starting from a relatively large  temperature of $T \sim 100 $ MeV,
the dependence of the IR term is basically governed by the Debye mass which suppresses
$B(\bp,\omega_n)$ at large $T$. This is a hint that in the considered models the Debye
mass has to be included only in the perturbative part of the interaction, i.e. in the
UV term only.}

\section{Impact of the IR term}\label{URimpact}
Analysing  the relative contributions of the IR and UV terms in the interaction we find that,
while at $T=0$  the UV term can be ignored in considering the hadron ground states~\cite{ourLast},
at finite temperatures the ultraviolet behavior can become important.

\begin{figure}[!ht]
\includegraphics[scale=0.6 ,angle=0]{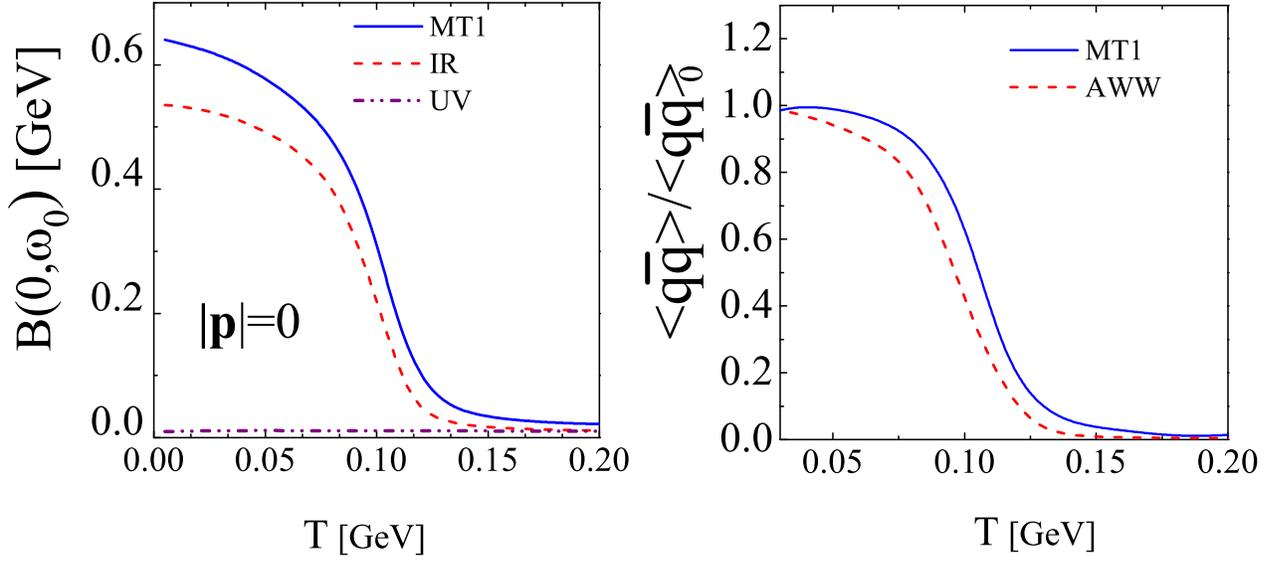}
\caption{(color online)
Relative contributions of IR and UV terms to the solution $B({\bp=0},\omega_0)$
of the tDS equation~(\ref{sde}) (left panel) and the quark condensate $\la q\bar q\ra$ (right panel)
 with IR term only (dashed curve), IR+UV terms (solid curve)
(right panel). In the left panel, the dashed and  dash-dot-dot curves represent the separate contributions
of   IR and UV terms, respectively. The solid curve is the total contribution of the IR+UV terms.
Effective parameters of the kernel are for the MT1 model.}
\label{fig8}
\end{figure}
  In Fig.~\ref{fig8}, left panel, we present the separate contributions of the IR (dashed curve) and
  UV (dash-dot-dot curve) terms  to the tDS solution $B({\bp=0},\omega_0)$. It is clearly seen that
  the absolute contribution of the UV term becomes sizeably  only at large temperatures.
  However, in the full kernel (solid curve,
  with both UV and IR term), the influence of UV part is  visible already  at low $T$.  This is because of
  interference effects and effects of higher Matsubara frequencies in the tDS equation~(\ref{sde}).
  {Nevertheless, the overall shape of $B({\bp=0},\omega_0)$ and, consequently the inflection
 point, is entirely governed by the IR interaction term. As mentioned above, all the
 considered interactions provide a  critical temperature significantly lower than the one in  lattice QCD
 calculations. Obviously, modifications of the kernel in the region above $T_c$ similar to   Eq.~(\ref{DotT})
 do not affect the behaviour of  $B({\bf 0},\omega_0)$ and $\la q\bar q\ra$ below such temperatures.
 Thus, a modification of the IR term in the whole range of $T$ is required.
 A possible modification of the kernel is as follows:
 (i) in the IR term the Debye mass is omitted,
 (ii) the parameter $D$ receives a $T$-dependence similar to  Eq. (\ref{DotT}),
 and (iii) the UV term, being inspired by perturbative QCD calculations, remains unchanged,
 i.e.}
 \begin{eqnarray} &&
 D^L(\bq,\Omega_{mn},m_g)=D_{IR}(\bq^2+\Omega_{mn})+D_{UV}(\bq^2+\Omega_{mn}^2+m_g^2),\\ \nonumber &&
 D_{IR}(k^2)=
        \frac{4\pi^2 D(T)  k^2}{\omega^6} e^{-k^2/\omega^2},
\ \quad
 D_{UV}(k^2+m_g^2) =
         \frac {8\pi^2 \gamma_m F(k^2+m_g^2)}{
            \ln[\tau+(1+\frac{k^2+m_g^2}{\Lambda_{QCD}^2})^2]} .
\label{modif}
\end{eqnarray}
 {
  The new effective
 parameters of the modified kernel should  smoothly approach their vacuum  values as $T$ approaches zero
  and must provide a suppression  of the IR interaction term   above the (pseudo-)
  critical temperature. As in the previous case above,
 a simple expression simulating such a behaviour may be written by utilizing
 two suppression functions with  a Heavyside step function-like behaviour,
 one acting below $T_c$ and, the second one acting above $T_c$, for example}

 \begin{eqnarray}
 D(T)=D\ \left[a \left\{1+\tanh\left(-\frac{T-T_p}{\beta}\right )\right\} +
 b \left\{1-\tanh\left (-\frac{T-T_p}{\beta'}\right)\right\}  \exp[{-\alpha^2 T^2]} \right],
 \label{newDott}
\end{eqnarray}
where $T_p$, $a$, $b$, $\alpha$, $\beta$ and $\beta'$ are new adjustable parameters.
In Fig.~\ref{fig9} we present an illustration of the change, according to Eq.~(\ref{newDott}),
of the IR parameter $D$,  computed with  a particular choice
of the effective parameters, $a=0.514$, $T_p=0.25$ GeV, $\beta=0.04$ GeV,
$b=5$, $\beta'=0.06$ GeV, and  $\alpha=10$ GeV$^{-1}$. At low temperatures,
$D$ is practically equal to its   vacuum value,
smoothly increases with temperature (making the interference with the UV term more pronounced),
 up to a maximum value ($\sim 6\%$) around
the  critical temperature, then it is screened at larger temperatures,
see also~\cite{robLast,kitaizyRpberts}.

\begin{figure}[!ht]
\includegraphics[scale=0.4 ,angle=0]{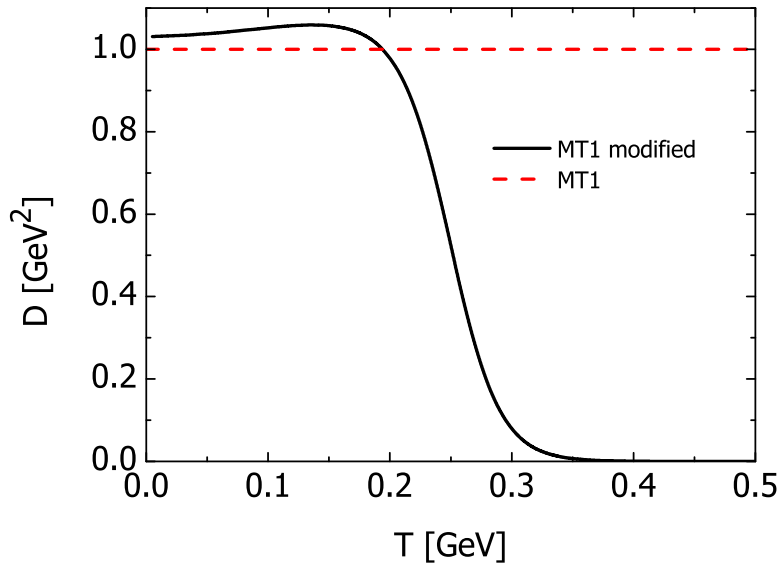}
\caption{(color online)
 A possible dependence of the  strength $D$ of the  IR term   in the MT1 and AWW  models on the
 temperature $T$ (solid curve) in comparison with the case of $D=const$ (dashed line).
The solution $B({\bp=0},\omega_0)$ and the quark condensate $\la q\bar q\ra$
with such a modified interaction are
exhibited  in Fig.~\ref{fig10}.}
\label{fig9}
\end{figure}
 The resulting solution $B({\bp=0},\omega_0)$,
as well as the quark condensate $\la q\bar q\ra$ for the modified interaction
are exhibited  in Fig.~\ref{fig10}, where the solid and dashed curves are for the
$T$-dependent solution for the MT1 and AWW models, respectively. It is
clearly seen that the inflection points are shifted towards  larger
values of temperature,   providing  critical temperatures $T_c\sim 135$ GeV for AWW and
$T_c\sim 140$ GeV for MT1, which now are better compatible with
the above quoted lattice values. This persuades us that the MT1 and AWW models with  proper modifications
 of the interaction kernel can provide a reasonable description of the quark propagators and quark
condensate at finite temperatures.
Such a modified interaction can be used then  in the BS equation to analyse the
behaviour of  mesons embedded in a hot environment.

\begin{figure}[!ht]
\includegraphics[scale=0.55 ,angle=0]{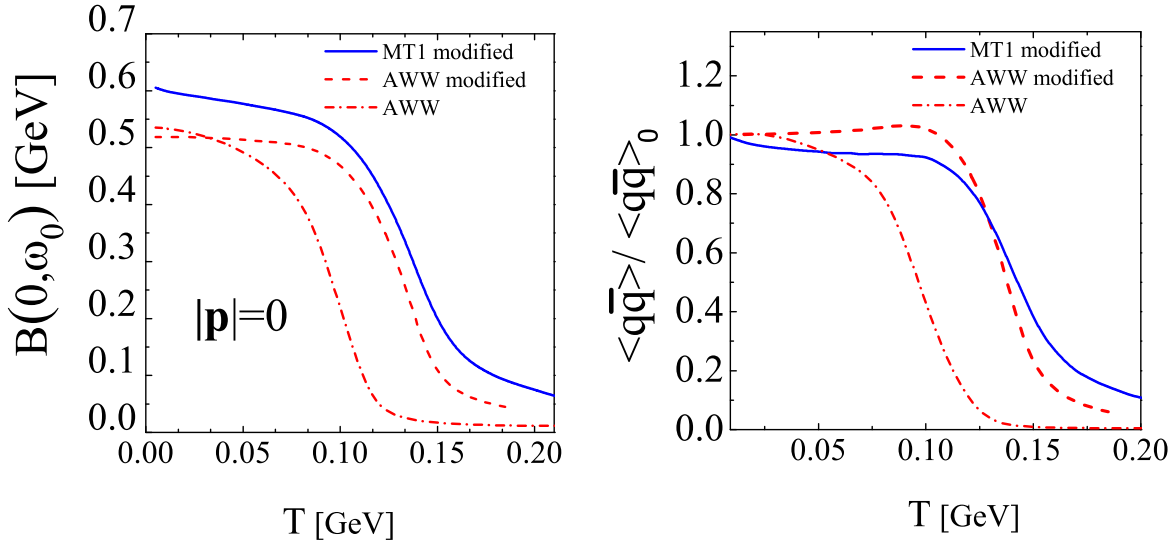}
\caption{(color online)
 The tDS solution $B({\bp=0},\omega_0)$ (left panel) and the quark condensate $\la q\bar q\ra$
 (right panel) obtained with the modified interaction, Eq. (\ref{newDott}). The solid and dashed
 curves represent results for the MT1 and AWW models, respectively.
The dot-dashed curve is for  the AWW solution  with $D=const$.  }
\label{fig10}
\end{figure}

Obviously, the effective parameters in Eq.~(\ref{newDott})
 can be   tuned further to obtain an improved  agreement with lattice calculations. This is not the goal of the
 present paper.
 We reiterate that we are interested in choosing an effective
 interaction suitable for solving the BS equation at finite temperature in a large interval of $T$, which can
 allow for  performing qualitative  analyses  of the behaviour of mesons in hot (and dense) matter as well as to infer  from this
 the relevant order parameters and other  conditions for a  possible phase transition at large temperature.
          \section{Summary}\label{summary}
 We have investigated  the impact of various choices of the
 effective quark-gluon interaction within the
 truncated rainbow approximations
 on the solution of the truncated Dyson-Schwinger  (tDS)
 equation  at finite temperature. The ultimate goal is to
 establish a reliable interaction kernel adequate in a large range
 of temperatures which, being used in  the Bethe-Salpeter equation,
 allows for an analysis of the behaviour of hadrons in hot matter, including possible phase transitions
 and  dissociation effects. For this we investigate to what extent the models, which provide an excellent
 description of mesons at zero temperatures, can be applied to the truncated tDS equation at finite temperatures.
  We find that in the chiral limit at temperatures below a critical value $T_c$ both  models, with and
  without the ultraviolet term, describe  fairly well the quark propagators.
  The critical temperature obtained from the condition of a zero
  of  the mass function $B$ and/or of the quark condensate is in agreement with
  calculations within the lattice or unquenched QCD. However, at temperatures above $T_c$
  the considered models with vacuum parameters fail to describe such important characteristics of the
  quark propagators as the quark spectral functions, thermal masses, plasmino mode etc. To achieve
   agreement of  the model calculations  with  QCD lattice results,
  a modification of the interaction kernels  is required. A simple change of the interaction is to suppress the contribution of the IR
  term at large temperatures, thus  making the interaction dependent on the temperature.
  At finite quark masses, the considered models seems to provide too small values for
  the (pseudo-) critical temperature which are  by ${\cal O}(50\%) $ smaller than
  the ones found in lattice QCD.  Modifications
  of the interaction in the same manner as for the chiral case, i.e. suppressing the IR term above
  $T_c$, do not affect the values of the (pseudo-) critical temperatures defined by  the
  maximum of the  susceptibility of the function $B$ or as the inflection point for
  the quark condensate. To obtain a larger value of the critical temperature
  the interaction kernel has to be modified also for smaller temperatures, even below $T\sim 100$ MeV.
  The Debye mass $m_g$ plays a crucial role in parametrizing the IR term. An inclusion of $m_g$
   as a Gaussian exponential, cf. Eq. (\ref{phenvf}),  results in an essential suppression
   of the solution of the tDS equation at $T > 100$ MeV,  making problematic the attempts
   of obtaining larger $T_c$, close to  the lattice values. It seems
   that the Debye mass ought to be included only in the perturbative part of the interaction.
   The $T$-dependence of the  IR term must be re-parametrized.
   The results of lattice calculations for the  $T$-dependence of the quark condensate suggest
   that  the $T$-dependence has  to be chosen  in such a way that at small temperatures
   the IR term approaches its vacuum value remaining  constant or smoothly changing
    up to   $T\sim 140 -150$ MeV;
   then it must be completely screened at larger temperatures.

  A more detailed parametrization of the IR term  requires a separate
  and   meticulous analysis of the tDS equation at finite $T$  and will
  be done elsewhere.
  \section*{Acknowledgments}
   This work was supported in part by the Heisenberg - Landau program
of the JINR - FRG collaboration, GSI-FE and BMBF. DSM and LPK appreciate the warm hospitality at the
Helmholtz Centre Dresden-Rossendorf.

                     \section{Appendix}

 \subsection {Rainbow truncation} The gap equation can be written as~\cite{MT}
\begin{equation}
S^{-1}(p) = Z_2  i\gamma   · p + Z_4 m_{bare}  +Z_1\int^\Lambda \frac{d^4 q}{(2\pi)^4}  g^2D_{\mu\nu}  (p -q)
\frac{\lambda_a}{2} \gamma_\mu S(q) \frac{\lambda_a}{2} \Gamma_\nu (q,p),
\label{full}
\end{equation}
where $D_{\mu\nu} $ is the dressed gluon propagator; $\Gamma_\nu$, the quark-gluon vertex;
$\int^\Lambda$   represents a Poincar\'{e} invariant regularization of the four-dimensional integral,
with $\Lambda$   the regularization mass-scale; $m_{bare}(\Lambda)$ denotes the current-quark bare
massand $Z_{i}  (\mu^2,\Lambda^2)$ stand for the corresponding  renormalisation constants,
with $\mu$  the renormalisation point,  and $\lambda_a$  is a Gell–Mann matrix acting
in color space.
  The solution of Eq. (\ref{full}) has the general form $ S(p)^{-1} = i\gamma \cdot ·p A(p^2,\mu^2) + B(p^2,\mu^2)$
 and is renormalized according
to $S(p)^{-1} = i \gamma \cdot p + m(\mu)$  at a sufficiently large  value of
 $ \mu^2$, with $m(\mu)$  the renormalized quark
mass at the scale $\mu$.     Since   the behaviour at high momenta $p^2 > 2$ GeV$^2$ is fixed by
 perturbation theory and the renormalisation flow,
 in concrete calculations on needs specify the  gap equation at low momenta, i.e. in the infra-red region.
 As   mentioned above,     in  studies of the quark DS equation one has to employ reliable model forms of the
 gluon propagator and quark-gluon vertex,  suitable
for the whole range of momentum squared $p^2$.
In rainbow-ladder  truncation, which is leading-order in the most widely used scheme, cf.
\cite{smekal,Alkofer,Maris:2003vk}, this is achieved by
adopting the requirements $Z_2=1$, $Z_4 m_{bare}=m$, where $ m$  is a
phenomenological  mass parameter,
$\Gamma_\nu (q,p)=\gamma_\nu$  and
\begin{equation}
Z_1g^2D_{\mu\nu} (k^2) =
  {\cal G}(k^2) D^{free}_{\mu\nu} (k^2)   =\left[D_{IR}  (k^2)  +
  4\pi \tilde\alpha_{QCD}(k^2)\right]  \left( \delta_{\mu\nu}-\frac {k_{\mu} k_{\nu}}{k^2} \right),
  \end{equation}
  where $˜\tilde\alpha_{QCD}(k^2)$  is a  smooth continuation of the perturbative-QCD running
coupling to all values of spacelike-$k^2 $  fulfilling the constraint of being finite  at the origin. The infra-red
term  $D_{IR}  (k^2) $  is constrained by the condition of the consistency with  Ward identities for the
tDS and tBS equations and to be negligibly small  in the perturbative region, i.e.
 $D_{IR}  (k^2) \ll  \alpha_{QCD}(k^2)$  at $k^2 \ge 2$ Gev$^2$.   Otherwise,    $D_{IR}  (k^2) $  is
 a pure phenomenological term, the form of which can be only qualitatively inferred from lattice calculations
 or from  solution  of    a (truncated) set of Dyson-Schwinger equations
  for the quark and gluon propagators within some additional reasonable approximations.  After
  choosing an explicit  form of the interaction, the  numerical values of the
  phenomenological  parameters  are determined from fitting empirical data.

  \end{document}